\documentclass[epj]{svjour}
\usepackage{graphics}
\begin{document}
\title{Comparing the reliability of networks by spectral analysis}
\author{Zitao Wang\inst{1} \thanks{\emph{Present address:} California Institute of Technology, 1200 East California Boulevard, Pasadena, CA 91125, USA} \and Kwok Yip Szeto\inst{2}
}                     
\institute{Department of Physics, The Hong Kong University of Science and Technology, Clear Water Bay, Hong Kong \and Department of Physics, The Hong Kong University of Science and Technology, Clear Water Bay, Hong Kong}
\date{Received: date / Revised version: date}
%
\abstract{
We provide a method for the ranking of the reliability of two networks with the same connectance. Our method is based on the Cheeger constant linking the topological property of a network with its spectrum. We first analyze a set of twisted rings with the same connectance and degree distribution, and obtain the ranking of their reliability using their eigenvalue gaps. The results are generalized to general networks using the method of rewiring. The success of our ranking method is verified numerically for the IEEE57, the Erd\H{o}s-R\'{e}nyi, and the Small-World networks.
} 
\maketitle
\section{Introduction}
\label{intro}

One of the early problems of statistical physics studied on networks are percolation processes \cite{Stauffer94}. Percolation model was first proposed and studied on real-world networks in the 1950s to model the spread of disease. In a percolation problem, one randomly designate each vertex or edge either ``open'' or ``closed'', and studies the various properties of the resulting patterns of vertices and edges \cite{Newman03,Cohen10,Newman10}. Here we are interested in the application of bond percolation to the question of network resilience.

For many biological, social, ecological and communication networks \cite{Albert02,Boccaletti06,Dorogovtesev03,Satorras04,Caldarelli07,Pierre95}, the issue of their resilience or fault tolerence to random failures of edges is of great importance as the dynamics taking place on these networks is affected by the changes in the corresponding network. For example, when some edge fails, the changes in the networks affect the interactions between the components of the complex systems and may lead to a complete failure of the entire system.  In engineering, fault tolerence is very important to cost minimization, while meeting certain acceptable level of service, as in the operation of communication systems \cite{Pierre95}, sewage systems and oil and gas lines \cite{Walters95}.  One obvious way to increase fault tolerence with extra cost is to introduce redundancy at the level of the components and connections. A balance between cost and fault tolerence can be achieved by design, through the search for an optimal topology of the network. 

Resilience to random failures of edges can be measured in different ways. Here we define a mathematical quantity called ``reliability'' to measure a network's resilience to random failures of edges. In this paper, we focus our discussion on undirected connected networks. Here a given network is said to be connected if every pair of vertices can be connected by at least one path (sequence of edges). We can then address the problem of reliability of the connected network using the well established framework of percolation theory.  Let us assume that every edge of the network are randomly designated either ``open'' (successful) or ``closed'' (failed). We denote by $p_R$ the probability that an edge is designated ``open'', and call it the \textit{bond open probability}. In engineering applications, this bond open probability is sometimes called the robustness of the edge. Now we can study the connectivity of the network subjected to random failure of the edges. We define the reliability of a given network to be the probability that the network remains connected given that the bond open probability is $p_R$. For example, when $p_R = 1$, the system reliability is $1$, because all edges are open with probability $1$, and the network is always connected when every edge works; whereas when $p_R=0$, the system reliability is $0$, because all edges closed (failed), and the network is always disconnected. The interesting question is the study of the reliability of a given system when $p_R$ is between $0$ and $1$.

The connectance of a network is the density of edges measured with respect to the fully connected network (complete graph). For undirected networks with $N$ vertices, the maximum number of edges is $L_{max}=N(N-1)/2$. The connectance $C$ is defined to be $L/L_{max}$. For a given undirected network with $N$ vertices and $L$ edges, the possible graphs with different topology is a large number, of the order of $L_{max} \choose L$. Among these graphs with the same connectance, we like to develop a simple mathematical theory for the comparison of their reliabilities. 

The topology of a network with the highest reliability is a fully connected network. However, as this topology requires the maximum number of edges, $L_{max}=N(N-1)/2$ for a network with N vertices,  the cost which is proportional to the number of edges is also maximum. Once we reduce the number of edges, the possible topologies exponentially increase, as there are $L_{max} \choose L$ number of topologies available for given $N$ and $L$, though not all of them are connected and cannot be used for the investigation of their reliability. Note that we always start with a connected network. Our objective is to compare the reliability of two connected networks with the same connectance (same $N$ and $L$) but different topology. 

\begin{figure}
\centering
\resizebox{0.4\textwidth}{!}{
\includegraphics{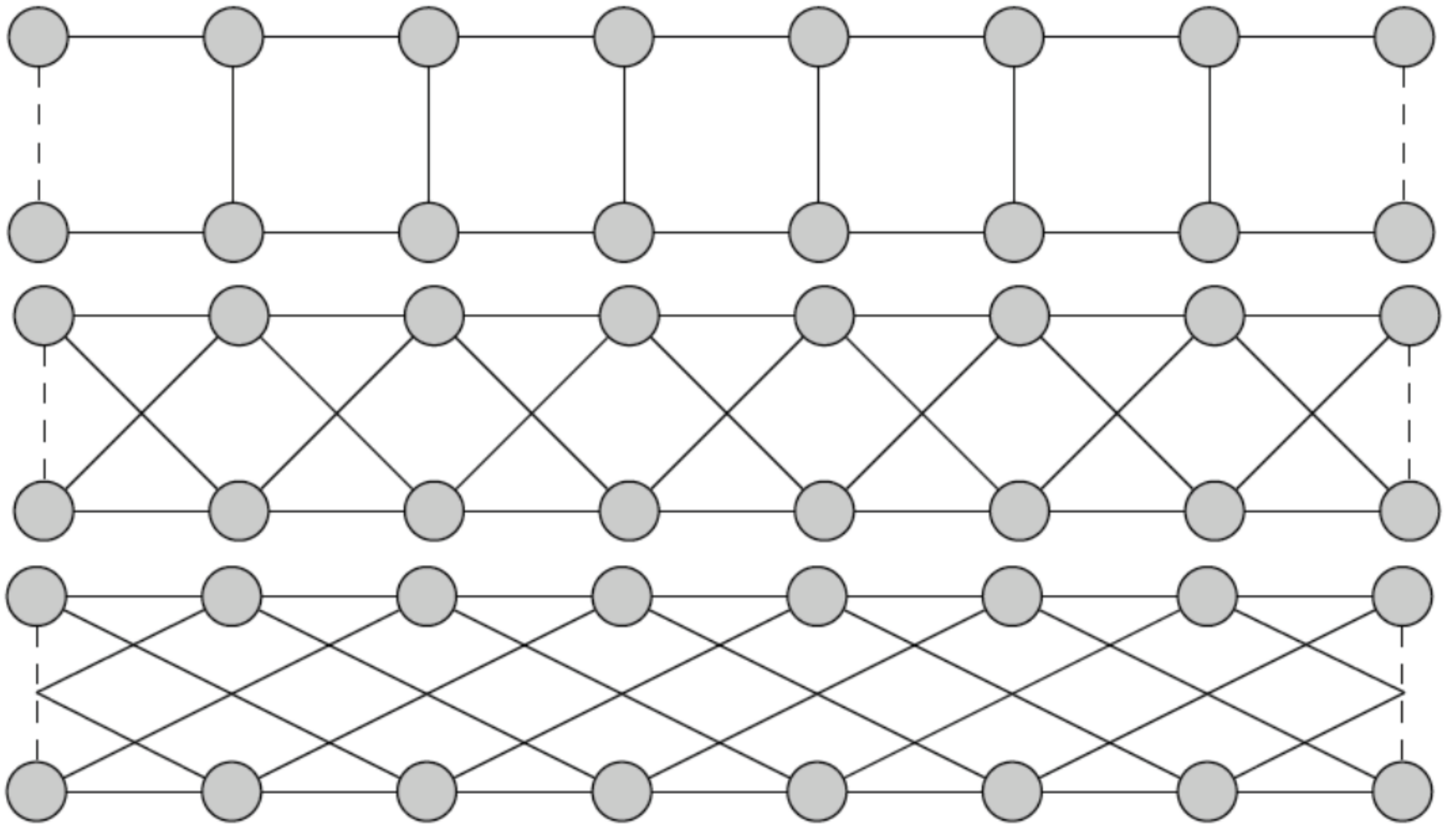}}
\caption{Examples of a “twisted ring” $R(M,N)$ topology of $N=8$ cells and $M=0$(top), $1$(middle), $2$(bottom) twist. The dashed lines refer to the periodic boundary condition to form the ring by identifying the vertices at the end of the band. The connectance of the $M=0$ topology is $1/5$ since $L_{max}=120$ and $L$ here is $24$, and that of the $M=1$ and $M=2$ topology is $4/15$ since $L_{max}$ is $120$ and $L=32$. \emph{Which topology is more reliable}?}
\end{figure} 

The comparison of the reliability of two real complex networks is difficult as they are often large. In order to obtain insight into the important features that determine the reliability of a network, we introduce some toy models which are sufficiently simple for mathematical analysis, and then generalize to large random networks with numerical verification of the conclusions obtained for the toy models. We first focus on some simple ring structures in the form of twisted rings $R(M,N)$ with $N$ unit cells and $M$ twists. In Fig.1 we show two twisted rings that have same number of vertices and edges, but are of different topologies. Each $N$-unit ring has $2N$ vertices and $3N$ or $4N$ edges (depending on $M$). A twist-$M$ ring has each vertex $k$ in the upper row connected to vertices in the lower row that are $M$-unit shifted with respect to $k$. The symmetry group of the ring networks is given by $C_2 \times D_{N}$, where $C_{2}$ is the cyclic group of order $2$, and $D_{N}$ is the dihedral group of order $2N$. The result is independent of the number of twists $M$ of the ring. Although these rings are sufficiently simple for mathematical analysis, they do reveal features that can be extended to the problem of reliability for general networks, including random ones. 

Our paper is organized as follows. In section 2, we compare the reliability of the twisted rings numerically and obtain a ranking order for several simple structures. In section 3, we introduce the usage of spectral gap, which can be computed exactly for these twisted rings, and make use of the relation between spectral gap and the Cheeger constant to obtain an explanation for the ordering of reliability of these twisted rings. In section 4, we give a more complex example using M\"obius strips formed by twisted rings to illustrate the power of our method of reliability comparison using spectral gap. In order to show that our method of spectral gap has application to real complex networks, we show that the spectral gap is very useful in comparing the reliability of larger real networks, such as the IEEE 57, Erd\H{o}s-R\'{e}nyi and Small-World networks. This is discussed in section 5, where the reliability of different ensembles of networks of different spectral gaps are compared for their reliability, while members in the ensemble are of the same connectance.   

\section{Reliability comparison of twisted rings}
\label{sec:2}

\begin{figure}
\centering
\resizebox{0.45\textwidth}{!}{
\includegraphics{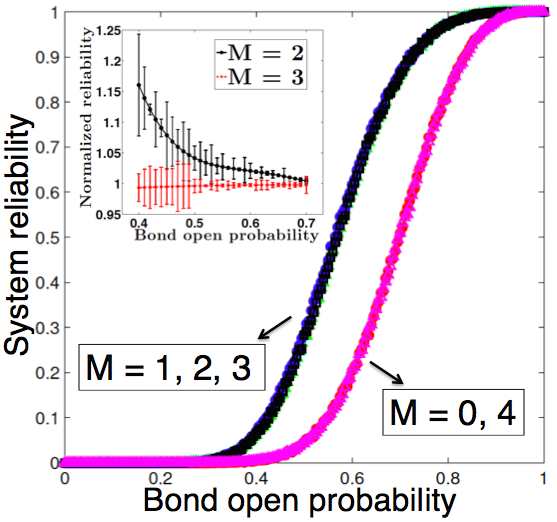}}
\caption{The general trend of the reliability of the network $R(M,N=8)$ versus the bond open probability for $M = 0,1,2,3,4$. The curves (in red lines) show the identical reliability for $R(0,8)$ with $R(4,8)$. The other curves (in black lines) are the reliability of $R(M,8)$ for $M = 1,2,3$. These curves are actually different and their differences can be shown in the inserted graph. The inserted graph shows the normalized reliability of $R(1,8)$, $R(2,8)$ and $R(3,8)$. Here the ratio of $R(2,8)$ and $R(3,8)$ over $R(1,8)$ is plotted against the bond open probability $p_R$. The error bar associated with the numerical estimate for reliability. We see that the ratio of $R(3,8)/R(1,8)$ is approximately $1$, which from our theory should be $1$ exactly, and that $R(2,8)>R(3,8)=R(1,8)$ for all $p_R$. Therefore, the answer to the question raised in Fig.1 is that $R(2,8)$ is the most reliable network among $R(M,8)$ for $M=1,2,3$.}
\label{fig.2}
\end{figure} 

For our rings, we use Monte Carlo simulation to compute numerically the probability of system working as a function of $p_R$. We first generate an ensemble of rings with the same topology (same $M$ and $N$). For each ring in the ensemble, we randomly (with probability $1-p_R$) cut its edges, so that the resultant network can be either connected or disconnected. We can then get a numerical value of the reliability of this topology by computing the ratio of number of networks that is connected in the ensemble over the total number of networks in the ensemble. The above process can be repeated for different topologies with different $M$ and $N$.

An example for the case of $8$-unit rings is shown in Fig.2. In order to distinguish the $M=1,2,3$ cases, we normalized the reliability curve of $R(2,8)$ and $R(3,8)$ by the reliability curve of $R(1,8)$.
The reliability ranking of all the 8-unit ring systems is given by
\begin{equation}
\label{eq:ranking}
R(2,8) \approx R(1,8) \approx R(3,8) > R(0,8) \approx R(4,8)
\end{equation}
We like to understand the reliability ranking presented in eq.(\ref{eq:ranking}). It is relatively easy to distinguish the reliability behavior of $R(M,8)$ ($M=1,2,3$) with that of $R(M,8)$ ($M=0,4$), because these two groups have different connectance, the former being $4/15$ whereas the latter being $1/5$. One can also see this from the largest eigenvalue of their respectively adjacency matrices, because for a regular network, its largest eigenvalue is equal to its degree. Hence a larger largest eigenvalue would imply a higher degree, hence higher connectance. Since systems with higher connectance have more redundancy for failure,$\ R(M,8)$ ($M=1,2,3$) would have higher reliability than $R(M,8)$ ($M=0,4$). This effect has been discussed in \cite{Masuda09,Restrepo08}.

\section{Spectral gap and the Cheeger constant for twisted rings }
\label{sec:3}

We like to understand the reliability ranking of $R(M,8)$ for $M=1,2,3$, i.e., group of networks with the same connectance. Can we predict which one is more reliable? It turns out that they are closely related to the gap between the largest eigenvalue and second largest eigenvalue of their respective adjacency matrices.

For the twisted ring $R(M,N)$, we can analytically compute its eigenvalue spectrum as a function of $M$ and $N$. The detailed calculation is shown in the appendix. We find that the eigenvalues of $R(0,N)$ are given by 
\begin{equation}
2\cos (2\pi i/N)\pm 1,0 \leq i \leq N-1.
\end{equation}
and the eigenvalues of $R(M,N)$($M > 0$) are given by
\begin{equation}
2\cos (2\pi i/N)\pm 2\cos (2\pi iM/N), 0 \leq i \leq N-1.
\end{equation}
when $N$ is odd and
\begin{equation}
2\cos (2\pi i/N)\pm 2\cos (2\pi iM/N) , \  0 \leq i \leq N-1, \ M\neq N/2. 
\end{equation}
\begin{equation}
2\cos (2\pi i/N)\pm (-1)^i ,\ 0 \leq i \leq N-1, \ M = N/2.
\end{equation}
when $N$ is even.

We compute the spectral gap of $R(M,8)$ (denoted as $G(M,8)$) using the above results and find that $G(M,8)$ are $0.5858$, $1.1716$, $2$, $1.1716$, $0.5858$ for $M=0,1,2,3,4$. The ranking of the gap is $G(2,8) > G(1,8) = G(3,8) > G(0,8) = G(4,8)$. In particular, for networks of the same connectance, their gap ranking is the same as their reliability ranking as shown in Fig.2.

The positive correlation between reliability and spectral gap for networks of the same connectance can be understood in terms of a mathematical quantity called the Cheeger constant\cite{Cheeger70}, named after the mathematician Jeff Cheeger. It is defined as follows: Let $X$ be an undirected network with vertex set $V(X)$ and edge set $E(X)$. One considers a partition of $V(X)$ into two disjoint subsets: $A \subseteq V(X)$ and its complement $B = V(X) \setminus A$. Denote by $\partial A \subseteq E(X)$ the set of edges connecting a vertex in $A$ and a vertex in $B$:
\begin{equation}
\partial A = \{(x,y) \in E(X) \ \vert x\in A, y\in B \}.
\end{equation}
The Cheeger constant of $X$, denoted by $h(X)$, is defined by 
\begin{equation}
h(X) = min\bigg\{\frac{ \vert \partial A \vert}{\vert A \vert} \ \bigg\vert A \subseteq V(X), 0 < \vert A \vert \leq \frac{\vert V(X) \vert}{2} \bigg\}.
\end{equation}
where $\vert X \vert $ is the cardinality of the set $X$, and the minimization is taken over all possible partitions of $V(X)$. 

By the above definition, the Cheeger constant has the following interesting properties\cite{Don06}: It is larger than zero if and only if the network is connected, and is ``large'' if any possible partition of the vertices has many edges between the two corresponding subsets. A physical interpretation of the Cheeger constant in the present context is that it is a measure of the size of the ``bottleneck'' ($\vert \partial A \vert$) connecting partition $A$ and $B$. A large Cheeger constant means a large bottleneck. In other words, the modularity of the network is poor. A small Cheeger constant means a small bottleneck, so that the network exhibits a very clear modular structure.

The Cheeger constant can be formally related to the spectral gap by the following inequality:
\begin{equation}
(d-\lambda)/2 \leq h(X) \leq \sqrt{2d(d-\lambda)}
\end{equation}
where $X$ is a finite, connected, $d$-regular graph, $\lambda$ is the second largest eigenvalue of its adjacency matrix, and $h(X)$ is the Cheeger constant of $X$. This inequality was proved by Dodziuk\cite{Dod84}, and independently by Alon-Milman\cite{AM85}, and Alon\cite{Alo86}.

As a consequence of the Cheeger inequality, upon increasing the spectral gap, the Cheeger constant of a network increases, and its modularity decreases, in the sense that it is more difficult to isolate subsets of vertices from the rest of the network. In other words, networks with larger spectral gap have topologies such that any set of vertices connects in a more robust way to all other vertices, which implies a higher reliability for such networks.

\begin{table*}
\caption{ Spectral gap and reliability of 7-unit rings and their respective M\"obius strips with $p_R = 0.5$.}
\label{tab.1}
\begin{center}
  \begin{tabular}{ c c c c c c c c c }
\hline
              & $R(0,7)$ & $Mb(0,7)$ & $R(3,7)$ & $R(1,7)$ & $Mb(1,7)$ & $Mb(2,7)$ & $Mb(3,7)$ & $R(2,7)$  \\ \hline
 Spectral gap & $0.7530$ & $0.7530$ & $0.9511$ & $1.5060$ & $1.5060$ & $1.7530$ & $2.3000$ & $2.3080$ \\ 
 Reliability & $0.0706$ & $0.0723$ & $0.2817$ & $0.3264$ & $0.3269$ & $0.3316$ & $0.3359$ & $0.3371$ \\
 Spectral gap ranking & $7$ & $7$ & $6$ & $4$ & $4$ & $3$ & $2$ & $1$ \\ 
 Reliability ranking & $7$ & $7$ & $6$ & $4$ & $4$ & $3$ & $2$ & $1$ \\ \hline
  \end{tabular}
\end{center}
\end{table*}  

As an example, we show how the above arguments work for the specific case of $8$-unit twisted rings. From the analytic computation of spectrum Eqs.(2)-(5), we obtain the spectral gap ranking of the $8$-unit rings: $G(2,8) > G(1,8) = G(3,8) > G(0,8) = G(4,8)$. Together with the Cheeger inequality, we can deduce the corresponding ranking of the Cheeger constants: $h(2,8) > h(1,8) = h(3,8) > h(0,8) = h(4,8)$. For small enough networks, it is not too complicated to calculate their Cheeger constants. In this case, 
\begin{equation}
h(2,8) = \frac{3}{2} > h(1,8) = h(3,8) = 1 > h(0,8) = h(4,8) = \frac{1}{2}
\end{equation}
which agrees with the spectral gap prediction. Now we have gathered enough information to make predictions about their reliability ranking. Since the connectance of $R(0,8)$ and $R(4,8)$ is $1/5$, whereas the connectance of $R(1,8)$, $R(2,8)$ and $R(3,8)$ is $4/15$. We expect the former group to have lower reliability. Inside each group of the same connectance, we compare their modularity by comparing their spectral gaps or Cheeger constants. Since larger spectral gaps implies larger Cheeger constants, which implies lower modularity and higher reliability, we expect the reliability ranking within each group to be: $R(0,8)=R(4,8)$; $R(2,8)>R(1,8)=R(3,8)$. So altogether we have $R(2,8) > R(1,8) = R(3,8) > R(0,8) = R(4,8)$, which agrees with numerical simulations Eq.(\ref{eq:ranking}). We should also emphasize that our spectral gap analysis is independent of the bond open probability $p_R$. This is also consistent with our numerical calculation shown in the insert of Fig.2 that the reliability ranking of $R(1,8)$, $R(2,8)$ and $R(3,8)$ is $R(2,8) > R(1,8) \approx R(3,8)$ for all $p_R$.   

In general, the calculation of topological quantities like the Cheeger constant for large networks is difficult, as the number of ways to partition the vertex set of a network with $N$ vertices is $2^{N-1}-1$, which grows exponentially with system size. Therefore, the computation of the Cheeger constant grows exponentially with system size. In comparison, most of the algorithms for eigenvalue computations scale polynomially with system size. Therefore, we expect that the algebraic criteria for comparing system reliability using spectral gap could be more powerful and efficient for large systems.

\section{Generalization via rewiring of Twisted ring to form M\"obius strip }
\label{sec:4}

The above conclusion can be applied to more general cases. One way to achieve this is to randomize a regular graph (do rewiring) to obtain the desired graph, and at the same time track the evolution of the spectral gap and reliability. We now start with the simplest example of rewiring on our twisted rings.

Let's do some rewirings of the twisted ring $R(M,N)$ to form M\"obius strips $Mb(M,N)$. We use $R(2,7)$ shown in Fig.3 as an example to illustrate how to construct M\"obius strips from the respective twisted rings. 

We first remove $6 = 2+2\times2$(and for general $R(M,N)$, $2M+2$) edges. For the particular labelling as in Fig.3, we first remove the two horizontal edges $01-07$ and $08-14$, so that the ring becomes a band. We then remove $01-13$ , $02-14$, $06-08$, $07-09$ (and for general R(M,N), these are a total of $2M$ edges), so that we can now do the twisting. We then twist one end of the band, say the right-hand side. Finally, we rewire back the removed edges, which are in this particular case, the following edges: $01-14$, $08-07$, $01-06$, $02-07$, $08-13$, $09-14$. The construction for general $Mb(M,N)$ is analogues to the above procedure with possibly more edges to be removed and glued back.

The number of vertices and edges and twists are unaffected. The spectral gap and reliability data for both the twisted rings and their respective M\"obius strips are summarized in Table 1, where we see the strict positive correlation between reliability and spectral gap are obeyed not only within the twisted rings and M\"obius strips, but also in the combined system. This is a significant extension of the original result, as rewirings break the original symmetry of the twisted-ring ensembles:  the original $C_2 \times D_{N}$ symmetry is no longer valid in the generalized ensemble, yet our conclusion still holds. Thus we may expect that the conclusion holds for general networks with similar connectance, since the heterogeneity in the vertex and edge distribution can always be smoothened by rewiring. A more challenging example is the rewiring of a given large network to form a Erd\H{o}s-R\'{e}nyi or a Small-World network. We find that the usage of the gap ordering for reliability still works, as shown in the next section.

\begin{figure}
\centering
\resizebox{0.4\textwidth}{!}{
\includegraphics{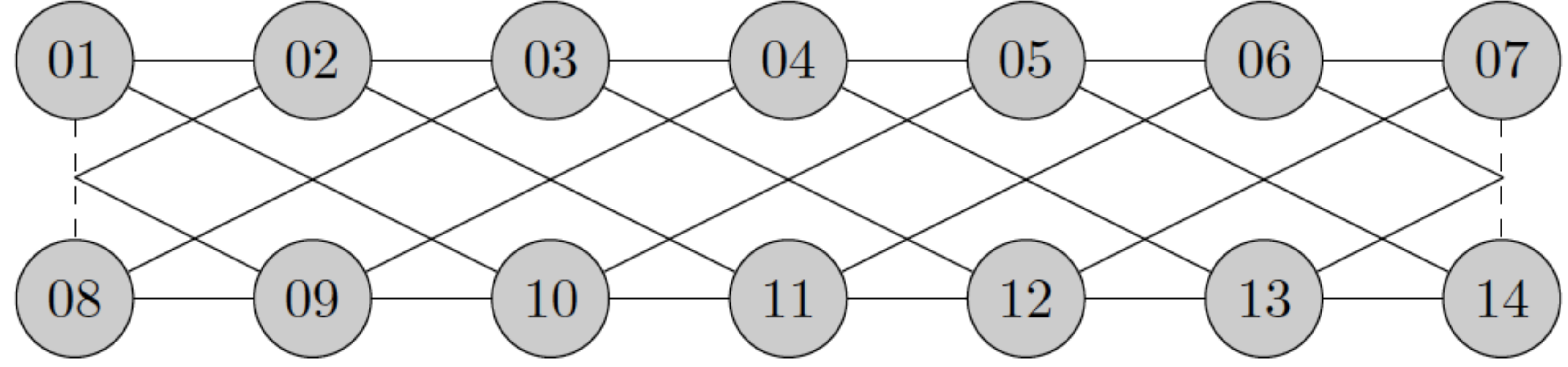}}
\caption{R(2,7)}
\label{fig.3}
\end{figure}

\section{Reliability studies of IEEE 57, Erd\H{o}s-R\'{e}nyi, and Small-World networks}
\label{sec:5}

\begin{table}
\caption{ Spectral gap and ensemble reliability with $p_R = 0.9$. The error bar shown in Table 2 are due to the number of samples, which is $100$,  in the ensemble of networks with the same connectance. There is no error in the IEEE 57 network as it is a well-defined network with $N=57$ nodes and $L=78$ edges. }
\label{tab.2}
\begin{center}
\resizebox{0.49\textwidth}{!}{
  \begin{tabular}{ c c c c } 
\hline
              & IEEE 57 & Erd\H{o}s-R\'{e}nyi & Small-World\\ \hline
 gap & $0.9430$ & $0.6273 \pm 0.1833$ & $0.4885\pm0.1441$ \\  
 reliability & $0.4720$ & $0.1684\pm0.0675$ & $0.1568\pm0.1166$\\ \hline
  \end{tabular}}
\end{center}
\end{table} 

We give a few more examples on the application of the spectral theory of reliability. We do a Monte Carlo experiment on the reliability of the IEEE 57\cite{Wang10}, Erd\H{o}s-R\'{e}nyi (ER)\cite{Erdos59,Erdos63}, and Small-World (SW)\cite{Kochen89,Watts99} Networks, and see how they are related to their respective spectra. The IEEE 57 network is used for the electric power grid of the Western United States and the New York state bulk electricity grid. To illustrate the relation between spectral gap and reliability of large networks, we perform rewiring without vertex removal on the IEEE 57 network to obtain new networks with the same connectance, but different topologies. We then compute the spectral gap and the reliability of these new networks and show that the ordering of spectral gap and the ordering of reliability remains positively correlated, as was shown for our toy models in sections 2-4. The construction of these new networks are illustrated in the following paragraph. They are adopted from the paper ``Statistical mechanics of complex networks" by R. Albert and A.-L. Barab\'asi\cite{Albert02}. Since we are interested in the effect of symmetry instead of connectance on network reliability, the free parameters in these experiments are controlled so that the three ensembles are of the same connectance.

\begin{enumerate}
\item
In the ER ensemble, each pair of vertices is connected with certain probability $p$. We choose $p$ so that the average connectance of the ensemble is the same as the connectance of the IEEE 57.

\item
In the SW ensemble, we start with a ring lattice with $L = 57$, and further connect vertex $1$ to vertex $3$, vertex $3$ to vertex $5$, and in general vertex $2k-1$ to vertex $2k+1$, until the network has the same connectance as IEEE 57. The resultant ``flower-shaped" network is well-ordered, of high clustering coefficient, and of high average path length.

\item
Randomly rewire each edge of the network in the SW ensemble with probability $p$ such that self-connections and duplicate edges are excluded. This process introduces long-range edges which connect vertices that otherwise would be part of different neighborhoods. By varying $p$ one can closely monitor the transition between order ($p = 0$) and randomness ($p = 1$). In order to facilitate comparison, we want the randomized network to be approximately of the same clustering coefficient as the IEEE 57 network. The clustering coefficient $C_{well-ordered}$ of the initial well-ordered network is $C(0) = C_{well-ordered} = 0.4971$, and we want $C_{rewired}\approx C_{IEEE 57} = 0.1222$, from which we can determine the rewiring probability to be $p \approx 0.45$. 
\newline

The results of the Monte Carlo experiments are summarized in Table 2. Note that both the ER ensemble and the SW ensemble has 100 members, and the recorded values in the table are the ensemble average values. The results show a positive correlation between spectral gap and reliability.
\end{enumerate}

\section{Conclusion and discussion}
\label{sec:6}

From these general results on rewired networks, we find that the correlation between the gap and the ordering of reliability may be useful for the design of networks when the number of edges and vertices are given. It implies that when comparing two networks for their reliability, the first thing to check is the connectance. Networks with higher connectance would have a higher reliability. If they have the same connectance, then we have to check for the second order effect, namely the symmetry of the networks. The network with a larger gap corresponds to one with higher symmetry, and has higher reliability. 

This theoretical tool is quite powerful, especially for comparing two large networks designed with same connectance, when other analytical or numerical methods are much more time-consuming. It is also possible to use our theory in the design of networks with enhanced performance with proper rewiring without vertex removal \cite{Wat10}. This is especially useful for the improvement of the reliability of existing networks. 

\section*{}
We acknowledge useful discussion with Chun Kit Chan and Ho Tat Lam. We acknowledge Grant No.FSGRF13SC25.

\section*{Appendix A. Analytical calculation of the twisted-ring spectrum}
\setcounter{equation}{0}
 \renewcommand{\theequation}{A.\arabic{equation}}

In the appendix, we present the analytical calculation needed to obtain Eqs.(2)-(5) in the main text.

An $n \times n$ matrix is called circulant, and denoted by $circ(a_1,a_2,\cdots,a_n)$ if it is of the form

\begin{equation}
circ(a_1,a_2,\cdots,a_n) = 
\left( \begin{array}{ccccc}
a_1 & a_2 & a_3 &\cdots & a_n \\
a_n & a_1 & a_2 &\cdots & a_{n-1}\\
\vdots & \vdots & \vdots & \ddots & \vdots\\
a_2 & a_3 & a_4 & \cdots & a_1 
\end{array}\right)_{n \times n}
\end{equation}  

Denote $C_n^k = circ(\underbrace{0,\cdots,0}_{k \ times},1,0,\cdots,1,\underbrace{0,\cdots,0}_{k-1 \ times})$, with corresponding eigenvalues $\beta_{i}^{(k)}, 0 \leq i \leq n-1 $. The vector that generates $C_n^k$ has all its entries equal to zero except the $(k+1)$-th entry and $(n-k+1)$-th entry, with the value of both entries equal to 1. The special case is when $n$ is even and $k = n/2$, these two entries overlap with each other, and the vector has all its entries equal to zero except the $(n/2+1)$-th entry.

The eigenvalues and eigenvectors for general circulant matrices have been well studied\cite{HLW06}, from which we can compute the eigenvalues and eigenvectors for $C_n^k$:

When $n$ is odd,
\begin{equation}
\beta_{i}^{(k)} = 2\cos (2\pi ik/n) , 0 \leq i \leq n-1.
\end{equation}

When $n$ is even,
\begin{equation} 
\left\{ \begin{array}{ll}
 \beta_{i}^{(k)} = 2\cos (2\pi ik/n) , 0 \leq i \leq n-1, k\neq n/2  \\
 \beta_{i}^{(k)} =(-1)^i , 0 \leq i \leq n-1, k = n/2
       \end{array} \right.
\end{equation}

Now we are ready to solve for the eigenvalue spectrum of $R(M,N)$. We begin with the simple example of square rings, i.e., the $M = 0$ cases (Fig.1). The $2N \times 2N$ adjacency matrix of $R(0,N)$ is

\begin{equation}
A(R(0,N))=
\left( \begin{array}{cc}
C_N^1 & I_N \\
I_N & C_N^1
\end{array} \right)
\end{equation}
where $I_N$ is the $N \times N$ identity matrix.

We construct an eigenvector of $R(0,N)$ of the form $\vec{u_i} = (a_i\vec{w_i},\vec{w_i})^t$ where $a_i$ is a constant, and $\vec{w_i}$ is an eigenvector of $C_N^1$ corresponding to $\beta_{i}^{(1)}$. To be more specific, $\vec{w_{i}}=(1,\omega ^i,\cdots,\omega ^{(N-1)i})^t , 0 \leq i \leq N-1,$ where $\omega$ is the $n$th root of unity. 

With such construction for $\vec{u_i}$, we have the following eigenvalue equation for $\lambda$:

\begin{equation}
\left(\begin{array}{cc}
C_N^1 & I_N \\
I_N & C_N^1
\end{array}\right)
\left(\begin{array}{c}
a_{i}\vec{w_{i}} \\
\vec{w_{i}}
\end{array}\right)
= \lambda
\left(\begin{array}{c}
a_{i}\vec{w_{i}} \\
\vec{w_{i}}
\end{array}\right)
\end{equation}

One can show that the only possible values for $a_i$ in Eq.(A.5) are $+1$ and $-1$, because $(\pm\vec{w_i},\vec{w_i})^t$, $0 \leq i \leq N-1$ constitute a complete set of eigenvectors for the block matrix (A.4).

Plug the $a_{i}'s$ back into Eq.(A.5), we get Eq.(A.6):
\begin{equation}
\left(\begin{array}{cc}
C_N^1 & I_N \\
I_N & C_N^1
\end{array}\right)
\left(\begin{array}{c}
\vec{\pm w_{i}} \\
\vec{w_{i}}
\end{array}\right)
=(\beta_{i}^{(1)}\pm1)
\left(\begin{array}{c}
\vec{\pm w_{i}} \\
\vec{w_{i}}
\end{array}\right)
\end{equation}

Combining with Eq.(A.2), the eigenvalues of $R(0,N)$ are given by 
\begin{equation}
\beta_{i}^{(1)}\pm1 = 2\cos (2\pi i/N)\pm 1,0 \leq i \leq N-1.
\end{equation}

Now we solve for the eigenvalue spectrum of a general twisted ring $R(M,N)$ ($M > 0$). The $2N \times 2N$ adjacency matrix of $R(M,N)$ is 
\begin{equation}
A(R(M,N)) = 
\left(\begin{array}{cc}
C_N^1 & C_N^M \\
C_N^M & C_N^1
\end{array}\right)
\end{equation}

Since any two circulant matrices commute, they have a commom set of eigenvectors $\vec{w_{i}}=(1,\omega ^i,\cdots,\omega ^{(N-1)i})^t , 0 \leq i \leq N-1,$. Therefore, similar to the $R(0,N)$ case, we can construct an eigenvector of $R(M,N)$ of the form $\vec{u_i} = (a_i\vec{w_i},\vec{w_i})^t$ where $a_i$ is a constant, and $\vec{w_i}$ is an eigenvector of both $C_N^1$ and $C_N^M$ corresponding to $\beta_{i}^{(1)}$ and $\beta_{i}^{(M)}$ respectively.

With such construction for $\vec{u_i}$, we have the following eigenvalue equation for $\lambda$:

\begin{equation}
\left(\begin{array}{cc}
C_N^1 & C_N^M \\
C_N^M & C_N^1
\end{array}\right)
\left(\begin{array}{c}
a_{i}\vec{w_{i}} \\
\vec{w_{i}}
\end{array}\right)
= \lambda
\left(\begin{array}{c}
a_{i}\vec{w_{i}} \\
\vec{w_{i}}
\end{array}\right)
\end{equation}

and similar calculation as in the $R(0,N)$ case yields $a_i = +1$ or $-1$, because $(\pm\vec{w_i},\vec{w_i})^t$, $0 \leq i \leq N-1$ constitute a complete set of eigenvectors for the block matrix (A.8).

Plug the $a_{i}'s$ back into Eq.(A.9), we get Eq.(A.10):
\begin{equation}
\left(\begin{array}{cc}
C_N^1 & C_N^M \\
C_N^M & C_N^1
\end{array}\right)
\left(\begin{array}{c}
\vec{\pm w_{i}} \\
\vec{w_{i}}
\end{array}\right)
=(\beta_{i}^{(1)}\pm \beta_{i}^{(k)})
\left(\begin{array}{c}
\vec{\pm w_{i}} \\
\vec{w_{i}}
\end{array}\right)
\end{equation}

Combining with Eq.(A.2) and Eq.(A.3), the eigenvalues of $R(M,N)$ ($M > 0$) are given by

When $N$ is odd,
\begin{eqnarray}
\beta_{i}^{(1)}\pm \beta_{i}^{(k)} 
= 2\cos (2\pi i/N)\pm 2\cos (2\pi iM/N), \nonumber \\ 0 \leq i \leq N-1. 
\end{eqnarray}

When $N$ is even,
\begin{eqnarray}
\beta_{i}^{(1)}\pm \beta_{i}^{(k)} 
= 2\cos (2\pi i/N)\pm 2\cos (2\pi iM/N), \nonumber \\  0 \leq i \leq N-1, \ M\neq N/2. 
\end{eqnarray}
\begin{eqnarray}
\beta_{i}^{(1)}\pm \beta_{i}^{(k)} 
= 2\cos (2\pi i/N)\pm (-1)^i, \ \ \ \ \ \ \ \ \ \ \ \ \ \nonumber \\ 0 \leq i \leq N-1, \ M = N/2.
\end{eqnarray}

\end{document}